\documentclass[12pt]{article}

\newcommand{\be}{\begin{equation}}
\newcommand{\ee}{\end{equation}}
\newcommand{\bea}{\begin{eqnarray}}
\newcommand{\eea}{\end{eqnarray}}
\newcommand{\bs}{\bigskip}

\newcommand{\s}{\smallskip}
\newcommand{\n}{\noindent}
\newcommand{\bc}{\begin{center}}
\newcommand{\ec}{\end{center}}
\newcommand{\bu}{\begin{underline}}
\newcommand{\eu}{\end{underline}}

\newcommand{\ty}{\textstyle}
\begin{document}

\bc{\Large {\bf A Gauge-Invariant Color Charge in QCD}}\ec \bs 
\bc{ Kurt Haller\footnote{E-mail: khaller@uconnvm.uconn.edu}\\
Department of Physics, University of Connecticut\\ Storrs, Connecticut
06269}\ec \bs
\begin{abstract}
\n
A gauge-invariant color-charge operator is defined and related to an integral of the gauge-invariant 
chromoelectric field over a closed surface. We discuss the case of a surface all of whose points
are a macroscopic distance from a system of quarks and gluons which it entirely surrounds.
When this system of quarks and gluons forms a  hadron or an object composed of hadrons, 
 such as a nucleus, it is argued that the gauge-invariant color charge 
enclosed within this surface must vanish and
 the system of hadrons in the interior of the surface must be a color singlet. 

\end{abstract}
In QCD, we encounter the quark color charge density $j^a_0({\bf r})=g\psi^{\dagger}({\bf r})
{\ty \frac{\lambda^a}{2}}\psi({\bf r})$ and the total color charge density $J^a_0({\bf r})+j^a_0({\bf r})$,
which includes the gluon color charge density which, 
in the temporal ($A^a_0=0$) gauge, which we choose for this work, 
is $J^a_0({\bf r})=gf^{abc}A^b_i({\bf r})\Pi^c_i({\bf r})$, 
where $\Pi^c_i({\bf r})$ is the negative chromoelectric
field as well as the momentum canonically conjugate to $A^c_i({\bf r})$.\footnote{In this work we 
use nonrelativistic notation, in which all space-time indices are subscripted and designate contravariant 
components of contravariant quantities such as $A_i^a$ or $j_i^a$, and covariant components of covariant 
quantities such as $\partial_i$.}  
The commutation rules that apply to 
these fields are
\begin{equation} [A^a_i({\bf x}), \Pi^b_j({\bf y})] =
i\delta_{ab}\delta_{ij}\delta({\bf x-y})\,;
\label{eq:comAiPjweylQ} 
\end{equation}
$j^a_0({\bf r})$ is not part of a conserved current.
It obeys the equation $D_0j^a_0+D_ij^a_i=0$ where $D_0j^a_0=\partial_0j^a_0$ and
$D_ij^a_i=(\partial_i\delta_{ab}+gf^{aqb}A^q_i)j^b_i$, with $j^a_i=g\psi^{\dagger}({\bf r})
{\ty \frac{\lambda^a}{2}}\alpha_i\psi({\bf r})$;  $j^a_0({\bf r})$ also is not gauge invariant --- it transforms like 
a vector in the adjoint representation of SU(3). In contrast, 
$J^a_0({\bf r})+j^a_0({\bf r})$ is a component of a conserved 
current. Together with $J^a_i+j^a_i$, where $J^a_i=gf^{abc}A^b_jF^c_{ij}$ and
$F^a_{ij}=\partial_jA^a_i-\partial_iA^a_j-gf^{abc}A^b_iA^c_j$, it forms the conserved current 
$\partial_0(J^a_0+j^a_0)+\partial_i(J^a_i+j^a_i)=0$. However, it is easily seen that the color charge 
$Q^a={\int}d{\bf r}\{J^a_0({\bf r})+j^a_0({\bf r})\}$ also is not gauge invariant. Since physical quantities 
must be gauge-invariant, it would be desirable to have a gauge-invariant color charge available in QCD.
In this paper, we will propose such a charge. \s

Elsewhere, we have constructed gauge-invariant operator-valued quark and gluon fields;\cite{CBH2} 
these include the gauge-invariant quark field
\begin{equation}
{\psi}_{\sf GI}({\bf{r}})=V_{\cal{C}}({\bf{r}})\,\psi ({\bf{r}})
\;\;\;\mbox{\small and}\;\;\;
{\psi}_{\sf GI}^\dagger({\bf{r}})=
\psi^\dagger({\bf{r}})\,V_{\cal{C}}^{-1}({\bf{r}})\;,
\label{eq:psiqcdg1}
\end{equation}
where
\begin{equation}
V_{\cal{C}}({\bf{r}})=
\exp\left(\,-ig{\overline{{\cal{Y}}^\alpha}}({\bf{r}})
{\textstyle\frac{\lambda^\alpha}{2}}\,\right)\,
\exp\left(-ig{\cal X}^\alpha({\bf{r}})
{\textstyle\frac{\lambda^\alpha}{2}}\right)\;,
\label{eq:el1}
\end{equation}
and
\begin{equation}
V_{\cal{C}}^{-1}({\bf{r}})=
\exp\left(ig{\cal X}^\alpha({\bf{r}})
{\textstyle\frac{\lambda^\alpha}{2}}\right)\,
\exp\left(\,ig{\overline{{\cal{Y}}^\alpha}}({\bf{r}})
{\textstyle\frac{\lambda^\alpha}{2}}\,\right)\;.
\label{eq:eldagq1}
\end{equation}   
In these expressions ${\cal{X}}^\alpha({\bf{r}}) {\equiv}
[\,{\textstyle\frac{\partial_i}{\partial^2}}A_i^\alpha({\bf{r}})]$, so that 
$\partial_i{\cal{X}}^\alpha({\bf{r}})$ is the $i$-th component of the longitudinal gauge field; and 
$\overline{{\cal Y}^{\alpha}}({\bf r}){\equiv}{\textstyle \frac{\partial_{j}}{\partial^{2}}
\overline{{\cal A}_{j}^{\alpha}}({\bf r})}$.
$\overline{{\cal A}_{j}^{\alpha}}({\bf r})$, which we 
refer to as the ``resolvent field'', is an operator-valued functional 
of the gauge field, and is represented in Refs.~\cite{CBH2} and \cite{HCC} as the 
solution of an integral equation that relates operator-valued quantities. 
Constructing a gauge-invariant quark field by 
attaching $V_{\cal{C}}({\bf{r}})$ to the quark field 
$\psi$ represents an extension, into
the non-Abelian domain, of a method of creating gauge-invariant charged fields 
originated by Dirac for QED;~\cite{diracgauge} and, like Dirac's procedure,
this non-Abelian construction is free of path-dependent integrals.
We have confirmed that ${\psi}_{\sf GI}({\bf{r}})$
is invariant to a non-Abelian gauge transformation, by showing that for a gauge 
transformation under which
\begin{equation}
{\psi}({\bf{r}})\,\rightarrow\,
\psi^\prime({\bf{r}})=\,
\exp\left(i\omega^\alpha({\bf{r}})\,
{\textstyle\frac{\lambda^\alpha}{2}}\,\right)\,\psi({\bf{r}})\;
\label{eq:psitransf}
\end{equation}
and 
\be
A^b_i({\bf{r}}){\textstyle\frac{\lambda^b}{2}}\,\,\rightarrow\,\exp\left(i\omega^\alpha({\bf{r}})
{\textstyle\frac{\lambda^\alpha}{2}}\right)
\left(A^b_i({\bf{r}}){\textstyle\frac{\lambda^b}{2}+\frac{i}{g}\partial_i}\right)
\exp\left(-i\omega^\alpha({\bf{r}})
{\textstyle\frac{\lambda^\alpha}{2}}\,\right)\,,
\ee
$V_{\cal{C}}({\bf{r}})$ also gauge-transforms as 
\begin{equation}
V_{\cal{C}}({\bf{r}})\rightarrow 
V_{\cal{C}}({\bf{r}})\exp\left(-i\omega^\alpha({\bf{r}})\,
{\textstyle\frac{\lambda^\alpha}{2}}\,\right)\;\;\;\;
\mbox{and}\;\;\;
V^{-1}_{\cal{C}}({\bf{r}})\rightarrow 
\exp\left(i\omega^\alpha({\bf{r}})\,
{\textstyle\frac{\lambda^\alpha}{2}}\,\right)
V_{\cal{C}}^{-1}({\bf{r}})\;  
\end{equation}
so that ${\psi}_{\sf GI}({\bf{r}})=V_{\cal{C}}({\bf{r}})\,\psi ({\bf{r}})$
remains gauge-invariant. The resolvent field also has an important role 
in the gauge-invariant gauge field,
\begin{equation}
A_{{\sf GI}\,i}({\bf{r}})=[\,A_{{\sf GI}\,i}^{b}({\bf{r}})\,{\textstyle\frac{\lambda^b}{2}}\,]
=V_{\cal{C}}({\bf{r}})\,[\,A_{i}^b({\bf{r}})\,
{\textstyle\frac{\lambda^b}{2}}\,]\,
V_{\cal{C}}^{-1}({\bf{r}})
+{\textstyle\frac{i}{g}}\,V_{\cal{C}}({\bf{r}})\,
\partial_{i}V_{\cal{C}}^{-1}({\bf{r}})\;,
\label{eq:AdressedAxz}
\end{equation}
which can be shown to be~\cite{CBH2}
\begin{equation}
A_{{\sf GI}\,i}^b({\bf{r}})=
A_{T\,i}^b({\bf{r}}) +
[\delta_{ij}-{\textstyle\frac{\partial_{i}\partial_j}
{\partial^2}}]\overline{{\cal{A}}^b_{j}}({\bf{r}})\;.
\label{eq:Adressedthree1b}
\end{equation}
We have also defined a gauge-invariant negative chromoelectric field,~\cite{BCH3,HGrib} by 
using earlier results to observe from Eq. (\ref{eq:AdressedAxz}) 
that, since in the temporal gauge $A_0=0$, 
\be
A_{\sf GI\;0}={\ty -\frac{i}{g}}V_{\cal{C}}\partial_0V_{\cal{C}}^{-1}\,.
\label{eq:pg2}
\ee
We recall an operator-order we have previously used,~\cite{CBH2}
and impose it on the definition of a gauge-invariant momentum 
(and negative gauge-invariant chromoelectric  field)
\be
{\Pi}_{{\sf GI}\,i}=-E_{{\sf GI}\,i}=||\partial_iA_{\sf GI\;0}+
\partial_0A_{\sf GI\,i}-ig\left[A_{\sf GI\,i},A_{\sf GI\;0}\right]||\,,
\label{eq:pg3}
\ee
where, using a notation introduced in Ref.\cite{CBH2}, 
bracketing between double bars denotes a normal order in which all gauge 
fields and functionals of gauge fields 
appear to the left of all momenta conjugate to gauge fields, but where
 that order is imposed only after all indicated 
commutators have been evaluated (including the commutator implied by the derivatives 
$\partial_0$ and $\partial_i$). From
\be
||\partial_0A_{\sf GI\,i}||=||\left[V_{\cal{C}}A_iV_{\cal{C}}^{-1}\,,
V_{\cal{C}}\partial_0V_{\cal{C}}^{-1}\right]+V_{\cal{C}}\partial_0A_iV_{\cal{C}}^{-1}
+\frac{i}{g}\left(\partial_0V_{\cal{C}}\partial_iV_{\cal{C}}^{-1}+
V_{\cal{C}}\partial_0\partial_iV_{\cal{C}}^{-1}\right)||
\label{eq:pg4}
\ee
and
\be
||\partial_iA_{\sf GI\;0}||=-||\frac{i}{g}\left(\partial_iV_{\cal{C}}\partial_0V_{\cal{C}}^{-1}+
V_{\cal{C}}\partial_0\partial_iV_{\cal{C}}^{-1}\right)||\,,
\label{eq:pg5}
\ee
combined with
\be
-ig\left[A_{\sf GI\,i},A_{\sf GI\;0}\right]=
-\left[V_{\cal{C}}A_iV_{\cal{C}}^{-1},V_{\cal{C}}\partial_0V_{\cal{C}}^{-1}\right]
-\frac{i}{g}\left[\partial_0V_{\cal{C}},\partial_iV_{\cal{C}}^{-1}\right]\,,
\label{eq:pg6}
\ee
we find that 
\be
{\Pi}_{{\sf GI}\,i}=||V_{\cal{C}}\partial_0A_iV_{\cal{C}}^{-1}||=||V_{\cal{C}}{\Pi}_{i}V_{\cal{C}}^{-1}||=
V_{\cal{C}}\textstyle{\frac{\lambda^b}{2}}V_{\cal{C}}^{-1}{\Pi}_{i}^b\,.
\label{eq:pg7}
\ee
By defining 
\be
{\cal R}_{db}={\textstyle\frac{1}{2}}{\sf Tr}[\lambda^dV_{\cal{C}}\lambda^bV_{\cal{C}}^{-1}],
\label{eq:R}
\ee 
we can express Eq. (\ref{eq:pg7}) in the equivalent form
\be
{\Pi}^d_{{\sf GI}\,i}={\cal R}_{db}{\Pi}_i^b\,.
\ee
It is similarly possible to show that \be
j^d_{\sf GI\;0}=g\psi^{\dagger}\,V_{\cal{C}}^{-1}
{\ty \frac{\lambda^d}{2}}\,V_{\cal{C}}\psi=R_{db}j^b_0\,,
\ee
and
\be
A_{{\sf GI}\,i}^d={\cal R}_{db}\,A_{i}^b-{\ty \frac{1}{g}}P_{di}
\ee 
where $P_{di}=-i{\sf Tr}[\lambda^d\,V_{\cal{C}}\,\partial_iV_{\cal{C}}^{-1}]$. 
The fields, $A_{{\sf GI}\,i}^d$, ${\Pi}^d_{{\sf GI}\,i}$ and ${\psi}_{\sf GI}({\bf{r}})$ are gauge-invariant, 
even though they carry color indices. All of them commute with the so-called ``Gauss's law operator'',
${\cal G}^a({\bf r})\equiv\partial_i\Pi^a_i({\bf r})+gf^{abc}A^b_i({\bf r})\Pi^c_i({\bf r})+j^a_0({\bf r})$.\s

We propose the following as the gauge-invariant color charge operator;
\be
{\cal Q}^a={\int}d{\bf r}\left\{gf^{abc}A^b_{{\sf GI}\,i}({\bf r})\Pi^c_{{\sf GI}\,i}({\bf r})+
j^a_{{\sf GI}\;0}({\bf r})\right\}\,.
\ee
By making the appropriate substitutions, we obtain
\be
{\cal Q}^a={\int}d{\bf r}\left\{f^{abc}\left(gR_{bp}R_{cq}A^p_i\Pi^q_i-P_{bi}R_{cq}\Pi^q_i\right)
+R_{ab}j^b_{0}\right\}\,.
\label{eq:gicharge}
\ee
From the identity
\be
f^{abc}({\sf T}^b)_{ij}({\sf T}^c)_{kp}={\ty \frac{i}{2}}\left(\delta_{jk}({\sf T}^a)_{ip}-
\delta_{ip}({\sf T}^a)_{kj}\right)\,,
\label{eq:epsSU2}
\ee
where ${\sf T}^a$ is a generator in a fundamental representation of an SU(N) group,
we obtain 
\be
f^{abc}R_{bp}R_{cq}={\ty \frac{i}{4}}{\sf Tr}\left\{\lambda^aV_{\cal{C}}\left[\lambda^q\,,
\lambda^p\right]V_{\cal{C}}^{-1}\right\}=f^{dpq}R_{ad}
\ee
and
\be
f^{abc}P_{bi}R_{cq}={\ty \frac{1}{2}}{\sf Tr}\left\{V_{\cal{C}}\lambda^c\partial_iV_{\cal{C}}^{-1}\lambda^a+
\partial_iV_{\cal{C}}\lambda^cV_{\cal{C}}^{-1}\lambda^a\right\}=\partial_iR_{aq}\,.
\ee
With these identities, Eq. (\ref{eq:gicharge}) becomes 
\be
{\cal Q}^a={\int}d{\bf r}\left\{R_{ab}\left[gf^{bpq}A^p_i\Pi^q_i+j^b_{0}\right]-(\partial_iR_{ab})\Pi^b_i\right\}
\label{eq:Q1}
\ee
which can also be expressed as
\be 
{\cal Q}^a={\int}d{\bf r}\left\{R_{ab}\left[\partial_i\Pi^b_i+gf^{bpq}A^p_i\Pi^q_i+j^b_{0}\right]-
\partial_i(R_{ab}\Pi^b_i)\right\}\,.
\label{eq:Q2}
\ee
When Eq. (\ref{eq:Q2}) is applied to a system consisting of quarks and gluons, which we represent as 
the state $|\Psi\rangle$,  
$\left[\partial_i\Pi^b_i+gf^{bpq}A^p_i\Pi^q_i+j^b_{0}\right]|\Psi\rangle=0$ must hold because it 
is required for the implementation of
Gauss's law or, equivalently, for the gauge invariance of $|\Psi\rangle$. This leads to 
\be
{\cal Q}^a\approx-{\int}d{\bf r}\left\{\partial_i(R_{ab}\Pi^b_i)\right\}\approx
-{\int}d{\bf r}\left\{\partial_i\Pi^a_{{\sf GI}\;i}\right\}
\label{eq:Q3}
\ee
where $\approx$ indicates a ``soft'' equality --- {\em i. e.} one that is only valid when the 
operators act on a state $|\Psi\rangle$.
Eq. (\ref{eq:Q3}) can be transformed into 
\be
\left({\cal Q}^a+{\int}d{\cal S}\left\{n_i\,\Pi^a_{{\sf GI}\;i}\right\}\right)|\Psi\rangle=0
\label{eq:Q4}
\ee
where ${\int}d{\cal S}$ represents integration over a surface enclosing the system of quarks and gluons
represented by $|\Psi\rangle$,  $n_i$ 
designates a unit outward normal to that surface, and the volume over which ${\cal Q}^a$ is 
integrated extends to the surface over which the surface integral is evaluated.\s

It is instructive to compare Eq.~(\ref{eq:Q4})
to the corresponding equation for QED. We find that this latter equation, for the electric charge, is 
\be
Q={\int}d{\bf r}\left\{\left[\partial_i\Pi_i+j_0\right]-\partial_i\Pi_i\right\}
\;\;\;\mbox{where}\;\;\;j_0=e\psi^\dagger\psi\;\;\mbox{and}\;\;\Pi_i=-E_i\,,
\label{eq:QED1}
\ee
which leads to 
\be
\left(Q+{\int}d{\cal S}\left\{n_i\,\Pi_i\right\}\right)|\Psi_{QED}\rangle=0\,,
\label{eq:QED2}
\ee
since $(\partial_i\Pi_i+j_0)|\Psi_{QED}\rangle=0$ expresses the Abelian Gauss's law 
and the requirement of gauge invariance that apply in this case.
In QED, Eq. (\ref{eq:QED2}) is the statement of the well-known result 
that the surface integral of the electric field $E_i=-\Pi_i$
over a closed surface is equal to the electric charge in the enclosed volume. 
If every point on the surface surrounding the charge is far enough from it --- a macroscopic distance, 
for example, from a charged particle or an ion --- the surface integral of $\Pi_i$ will not be affected by
quantum fluctuations of the charge density, and the classical result will be obtained.
It might seem possible to make a similar argument 
about QCD, based on the equation
\be
Q^a={\int}d{\bf r}\left\{gf^{abc}A^b_i({\bf r})\Pi^c_i({\bf r})+
j^a_0({\bf r})\right\}
\label{eq:QF1}
\ee
from which we obtain
\be
\left(Q^a+{\int}d{\cal S}\left\{n_i\,\Pi^a_i\right\}\right)|\Psi\rangle\,=0.
\label{eq:QF2}
\ee
However, Eq. (\ref{eq:QF2}) does not permit us to draw firm conclusions about $Q^a$. 
$\Pi^a_i$ and $Q^a$ are gauge-dependent, and therefore cannot refer to physical quantities. A finite SU(3)
gauge transformation is complicated to represent, but for the SU(2) case, the effect of a 
gauge transformation on $\Pi^a_i$ by a finite gauge function $\omega^a({\bf r})$ is 
\be
\left(\Pi^a_i\right)^\prime=\Pi^a_i-\epsilon^{abc}\omega^b\,\Pi_{i}^c\,
{\textstyle\frac{\sin(|\omega|)}{|\omega|}}\,-\epsilon^{aqd}\epsilon^{bcq}\,
\omega^b\omega^d\,\Pi^c_{i}\,
\,{\textstyle\frac{1-\cos(|\omega|)}{|\omega|^2}}\,.
\ee
The surface integral given in Eq. (\ref{eq:QF2}) is subject to arbitrary and 
potentially sizable changes when a 
gauge transformation is made, and no reliable inference about an 
observable color charge can be based on this equation.\s

With Eqs.~(\ref{eq:Q2}) - (\ref{eq:Q4}), we have provided the same kind of 
relation between color charges and the chromoelectric field that Eqs.~(\ref{eq:QED1}) and (\ref{eq:QED2}) provide
about the electric charge and the electric field in QED --- relations between gauge-invariant charge operators and integrals
over gauge-invariant fields. We are therefore in a position to make an argument similar 
to the  one applied to QED: that the chromoelectric field (like the electric field, in QED)
on a surface surrounding a system of quarks and gluons,
every point of which is far removed from the quarks and/or gluons it contains, is no longer 
subject to quantum fluctuations in its remote interior. It is, 
in fact, possible to draw an even stronger conclusion about the gauge-invariant color charge than about the 
electric charge. Since --- in contrast to the electric field ---
we can exclude the possibility, on physical grounds, that the chromoelectric field
extends over macroscopic distances from the hadrons or ensembles of hadrons that give rise to it, we 
can make the following argument: For a closed surface that surrounds a hadron or a system 
of hadrons such as a nucleus, all of whose points  ${\bf r}_s$ are at a macroscopic distance from the hadrons in 
its interior,  $\Pi^a_{{\sf GI}\;i}({\bf r}_s)|\Psi_{h}\rangle=0$, 
where $|\Psi_{h}\rangle=0$ represents a hadron or a system of hadrons
in the interior of the closed surface.  For such a surface surrounding this system of hadrons,
the surface integral 
\be
{\int}d{\cal S}\left\{n_i\,\Pi^a_{{\sf GI}\;i}({\bf r}_s)\right\}|\Psi_{h}\rangle=0\,.
\ee
Therefore, the total gauge-invariant color charge of the enclosed hadron or groups of hadrons 
must be given by
\be 
{\cal Q}^a|\Psi_{h}\rangle=0.
\ee
We can also use Eq. (\ref{eq:Q2}) to represent the commutator of two components of the gauge-invariant 
color charge operator as
\be
\left[{\cal Q}^a\,,{\cal Q}^b\right]=\left[\left({\int}d{\bf x}\,{\cal G}^a_{\sf GI}({\bf x})-\!\!
{\int}d{\cal S}_{\bf x}\left\{n_i\,\Pi^a_{{\sf GI}\;i}({\bf x})\right\}\right)\,,
\left({\int}d{\bf y}\,{\cal G}^b_{\sf GI}({\bf y})-\!\!
{\int}d{\cal S}_{\bf y}\left\{n_j\,\Pi^b_{{\sf GI}\;j}({\bf y})\right\}\right)\right]
\label{eq:comm1}
\ee
where
\be
{\cal G}^a_{\sf GI}=\partial_i\Pi^a_{{\sf GI}\;i}+
gf^{apq}A^p_{{\sf GI}\;i}\Pi^q_{{\sf GI}\;i}+j^a_{{\sf GI}\;0}=
R_{ab}\left(\partial_i\Pi^b_i+gf^{bpq}A^p_i\Pi^q_i+j^b_{0}\right).
\label{eq:comm2}
\ee
Since it has been shown that~\cite{HGrib}
\be
\left[{\cal G}^a_{\sf GI}({\bf x})\,,{\cal G}^b_{\sf GI}({\bf y})\right]=-igf^{abc}
{\cal G}^c_{\sf GI}({\bf x})\delta({\bf x}-{\bf y})\,,
\label{eq:comm3}
\ee
it follows that
\bea
&&\left[{\cal Q}^a\,,{\cal Q}^b\right]=-if^{abc}{\cal Q}^c+\left[{\int}d{\bf x}\,
{\cal G}^a_{\sf GI}({\bf x})\,,{\int}d{\cal S}_{\bf y}\left\{n_j\,\Pi^b_{{\sf GI}\;j}
({\bf y})\right\}\right]
-\left[{\int}d{\bf y}\,{\cal G}^b_{\sf GI}({\bf y})\right.\,,\nonumber\\&&\left.{\int}d{\cal S}_{\bf x}
\left\{n_i\,\Pi^a_{{\sf GI}\;i}({\bf x})\right\}\right]+\left[{\int}d{\cal S}_{\bf x}
\left\{n_i\,\Pi^a_{{\sf GI}\;i}({\bf x})\right\}\,,{\int}d{\cal S}_{\bf y}\left\{n_j\,\Pi^b_{{\sf GI}\;j}
({\bf y})\right\}\right]\,.
\label{eq:comm4}
\eea
In applying Eq. (\ref{eq:comm4}) to a hadron, the following remarks are germane:
the surface surrounding a hadron can be chosen with every point at a macroscopic 
distance from that hadron. In that case, the heuristic assumption we have already made ---
that the chromoelectric field at that surface vanishes --- leaves the commutator as
\be
\left[{\cal Q}^a\,,{\cal Q}^b\right]\approx-if^{abc}{\cal Q}^c\,,
\label{eq:comm5}
\ee
where  $\approx$ signifies that the equality only applies to the case in which the operators 
act on a state $|\Psi_{h}\rangle$. Eq. (\ref{eq:comm5}) is 
obviously consistent with ${\cal Q}^a|\Psi_{h}\rangle=0$
for all values of $a$. Furthermore, the time-derivative of ${\cal Q}^a$ can be represented as 
$i[{\hat H}_{\sf GI}\,,{\cal Q}^a]$ 
where ${\hat H}_{\sf GI}$ is the QCD Hamiltonian represented in terms of gauge-invariant fields.~\cite{HGrib}
From  Eq. (\ref{eq:Q2}) and from a result obtained in Ref.~\cite{HGrib},
\be
\left[{\hat H}_{\sf GI},\,{\cal G}^a_{\sf GI}({\bf{x}})\right]=
{\int}d{\bf r}{\chi}^{ac}({\bf{r}},{\bf{x}}){\cal G}^c_{\sf GI}({\bf r})
\label{eq:ggitime2}
\ee
where ${\chi}^{ac}({\bf{r}},{\bf{x}})$ is a nonlocal functional of gauge-invariant fields,
we obtain
\be
i\left[{\hat H}_{\sf GI}\,,{\cal Q}^a\right]|\Psi_{h}\rangle=i\left[{\hat H}_{\sf GI}\,,
{\int}d{\cal S}\left\{n_i\,\Pi^a_{{\sf GI}\;i}\right\}\right]|\Psi_{h}\rangle=0\,,
\ee
so that the gauge-invariant color charge of a hadron or system of hadrons is time-independent.\s

A state vector $|\Psi\rangle$ for which ${\cal Q}^a|\Psi\rangle=0$ for all $a$ has to be a color singlet. 
The preceding  arguments, based partly on formal properties of gauge-invariant operator-valued fields
and additional physical considerations, 
 lead us to the following conclusion: A hadron, or an ensemble of hadrons
contained in a nucleus or occupying 
some other well-defined microscopic region of space, has to be a color singlet.
An isolated quark or object composed of quarks and/or gluons, 
that is so far removed from other quarks and gluons that a 
surface can be drawn with every point a macroscopic distance from it,
 without enclosing any further gluons or quarks other than that isolated object 
in its remote interior, can have a 
nonvanishing gauge-invariant color charge only if it gives rise to a gauge-invariant chromoelectric 
field that persists up to macroscopic distances from that object. The fact that long-range 
chromoelectric fields are not observed, accounts for the fact that ensembles of quarks and gluons
 that are not color singlets cannot exist as isolated states 
\s

This research was supported by the Department of Energy under Grant
No.~DE-FG02-92ER40716.00.


\begin{thebibliography}{99}
\bibitem{CBH2}L. Chen, M. Belloni and K. Haller, Phys.\ Rev.
{\bf D 55} (1997) 2347.
\bibitem{HCC}K. Haller, L. Chen and Y. S. Choi, Phys. Rev. {\bf D 60} (1999) 125010.
\bibitem{diracgauge} P. A. M. Dirac, Canad. J. Phys. {\bf 33}, 650 (1955).
\bibitem{BCH3} M. Belloni, L. Chen and K. Haller, Phys.\ Lett.
{\bf B 403} (1997) 316.
\bibitem{HGrib}K. Haller, Int. Jour. Mod. Phys. {\bf A16} (2001) 2789.
\end{thebibliography}
\end{document}